\documentclass[reprint,amsmath,amssymb,aps,superscriptaddress]{revtex4-2}
\usepackage{graphicx}
\usepackage{dcolumn}
\usepackage{bm}

\begin{document}

\title{Evidence of topological gap opening in the surface state of Bi$_2$Se$_3$ by proximity to a magnetic insulator}

\author{S. Mathimalar}
\thanks{equal contribution}
\affiliation{Tata Institute of Fundamental Research, Centre for Interdisciplinary Sciences, Hyderabad-500107, India}

\author{S. Sasmal}
\thanks{equal contribution}
\affiliation{Tata Institute of Fundamental Research, Centre for Interdisciplinary Sciences, Hyderabad-500107, India}

\author{P. Rajasekhar}
\thanks{equal contribution}
\affiliation{Tata Institute of Fundamental Research, Centre for Interdisciplinary Sciences, Hyderabad-500107, India}

\author{A. Bhardwaj}
\affiliation{Tata Institute of Fundamental Research, Centre for Interdisciplinary Sciences, Hyderabad-500107, India}

\author{S. Chaudhary}
\affiliation{Tata Institute of Fundamental Research, Centre for Interdisciplinary Sciences, Hyderabad-500107, India}

\author{B. Satpati}
\affiliation{Saha Institute
of Nuclear Physics, 1/AF Bidhannagar, Kolkata, 700064, India}

\author{K. V. Raman}
\email[Corresponding author: ]{kvraman@tifrh.res.in}
\affiliation{Tata Institute of Fundamental Research, Centre for Interdisciplinary Sciences, Hyderabad-500107, India}

\begin{abstract}
Topological insulators are bulk insulators with exotic surface states, protected under time-reversal symmetry, that hold promise in observing many exciting condensed-matter phenomena. In this report, we show that by having a topological insulator (Bi$_2$Se$_3$) in proximity to a magnetic insulator (EuS), a metal-to-insulator transition in the surface state, attributed to opening of an exchange gap, can be observed whose properties are tunable using bottom gate voltage and external magnetic field. Our study provides evidence of gate-controlled enhanced interface magnetism with the signature of half-integer quantum Hall effect when the Fermi level is tuned into the exchange gap. These results pave the way for using magnetic proximity effect in developing topological electronic devices.

\end{abstract}
\maketitle
The process of breaking time reversal symmetry, by introducing out-of-plane magnetism, at the surface of a 3-dimensional (3-D) topological insulator (TI) \cite{hasan2010,qi2011} is known to open up an exchange gap (EG) at the Dirac point of the topologically protected surface states (TSS) \cite{chen2010}. Such a mechanism support several new quantum phenomena, such as quantum anomalous Hall effect (QAHE) \cite{yu2010,chang2013}, half-integer quantum Hall effect (H-IQHE) \cite{hasan2010,gusynin2005}, topological magneto-electric effect \cite{qi2010,essin2009}, and image magnetic monopoles \cite{qi2009}. In realizing these effects, the main challenge is to successfully demonstrate an opening of gap at the surface of TI.  Currently, this is made possible in magnetic TIs such as in Cr and V-doped compensated TIs, where perpendicular magnetism is developed in the bulk showing QAHE \cite{chang2013,mogi2017}. In comparison, inducing local magnetism at TI surface through the short range nature of magnetic proximity effect (MPE) can provide a number of advantages over bulk-doping \cite{chen2010,chang2013,hor2010}; these include lower bulk and surface defect density due to improved film crystallinity, better controllability of the surface states (SSs) and independently controlling the magnetic state at each surface of TI to build switchable topological devices \cite{Hou2019}.

\par
\begin{figure*}[t]
    \centering
    \includegraphics[width=\textwidth]{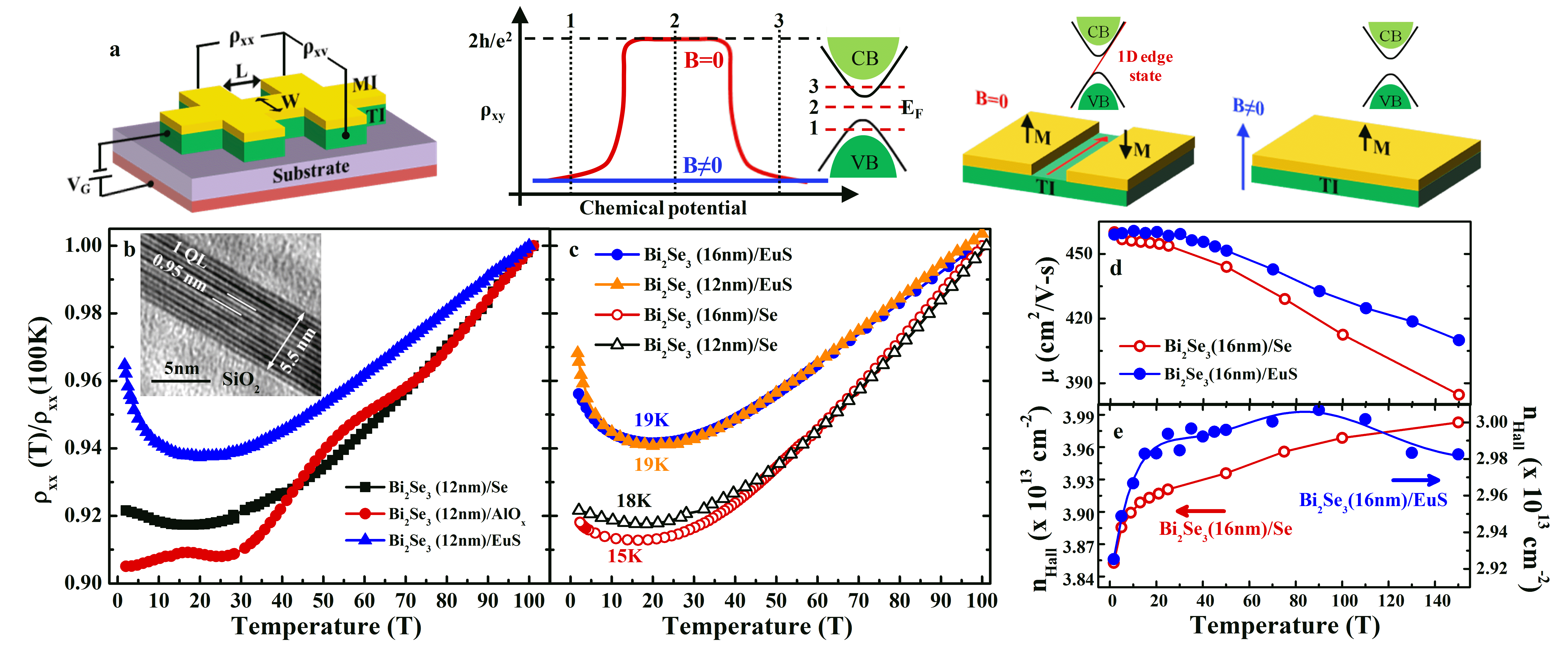}
    \caption{(a) Schematic of our TI/MI Hall bar devices (left) with bottom gate electrode. Chemical potential dependency of Hall resistivity ($\rho_{xy}$) in B=0 (red) and B$\neq$0 (blue) demonstrating H-IQHE is shown. Three different positions (1,2 and 3) of surface E$_F$, modulated by V$_g$, is labelled in the exchange gapped SSs of TI. For B=0, When E$_F$ lies in the EG (position 2), half-integer quantized value of $\rho_{xy}$ is expected which vanish when E$_F$ moves out of the EG (position 1 and 3). At B=0, 1D chiral Fermion edge states appear along magnetic domain walls which vanish in external magnetic field due to alignment of domains. (b) Normalized longitudinal resistivity ($\rho_{xx}$) vs $T$ of 12 nm Bi$_2$Se$_3$ films capped by 2 nm-AlO$_x$ (red circles), 2 nm-Se (black square) and 3 nm-EuS (blue triangle). Inset shows cross-sectional TEM image of sample: Si/SiO$_2$/Bi$_2$Se$_3$(5.5 nm). (c) Thickness dependence of normalize $\rho_{xx}$ vs T for the Se capped (hollow symbols) and EuS capped device (solid symbols) with 12nm (triangles) and 16nm (circles) thickness of Bi$_2$Se$_3$. Corresponding temperature value of the resistivity up-turn is also indicated. Temperature dependence of (d) $\mu$ and (e) $n_{Hall}$ for Bi$_2$Se$_3$(16nm)/Se and Bi$_2$Se$_3$(16nm)/EuS devices.}  
    \label{fig:1}
\end{figure*}

 Conventionally, MPE can be achieved by using magnetic insulators (MIs). However, requirement of an out-of-plane magnetization at the TI surface makes this study difficult since most of the MIs naturally possess in-plane anisotropy. Additionally, new hybridization states \cite{menshov2013} that may form at the TI/MI interface near the Fermi level (E$_F$) can contaminate the topological properties of the Dirac SSs \cite{luo2013,eremeev2013,eremeev2015}. As a result, observing these effects through transport studies can become non-trivial. In this regard, MPE studies in the 3D TI, Bismuth Selenide (Bi$_2$Se$_3$), with an MI, Europium sulphide (EuS), is extensively researched \cite{wei2013,katmis2016,yang2013}. Different research groups confirm the presence of an out-of-plane component of magnetization at the Bi$_2$Se$_3$/EuS interface with the magnetism existing at temperature much above the magnetic transition temperature ($T_c$) of EuS ($\sim17$K). Our magnetometry studies on Bi$_2$Se$_3$(10nm)/EuS(1nm) also confirm the existence of enhanced interface magnetism with a second transition close to 232K (refer to Fig. S8). Here, the strong spin-orbit contribution of the Bi$_2$Se$_3$ surface is reported to give rise to an out-of-plane magnetic anisotropy in the interface EuS layer \cite{katmis2016,kim2017}. Unlike other TI/MI interfaces \cite{luo2013,eremeev2013,lang2014,liu2015}, recent theoretical calculation of the above interface also suggest induced magnetism, localized within the first quintuple layer (QL) of Bi$_2$Se$_3$ \cite{lee2014}, with the SS retaining most of the topological character compared to a pristine Bi$_2$Se$_3$ SSs \cite{kim2017}. Further, an induced EG of $\sim9$ meV in the surface Dirac cone is suggested \cite{lee2014}. Despite these reports, experimental observation of the EG opening at this surface has been elusive \cite{krieger2019}. This is primarily due to inherent intrinsic doping caused by large Se vacancies leading to bulk dominated conduction with E$_F$ pinning deep inside the conduction band (CB). In this letter, using gate controlled transport studies, we provide evidence of gaining access to the exchange gapped TSSs at the Bi$_2$Se$_3$/EuS interface showing signature of H-IQHE (Fig.\ref{fig:1}a)  and observe a gate and magnetic-field controlled interface magnetism affecting the metal-to-insulator transition (MIT) response. 

\par Thin films of Bi$_2$Se$_3$ were grown on amorphous SiO$_2$ (a-SiO$_2$) substrate at 250$^{\circ}$C in our molecular beam epitaxy system with a base pressure of $8\times10^{-10}$ mbar using a co-evaporation technique. The film's structural properties were characterized using Raman spectroscopy, X-ray diffraction and Transmission electron microscopy (TEM) (see Fig. S1 in Suppl. Mater.). Interestingly, cross-sectional TEM images reveal a layered growth, supporting the hexagonal structure with 0.95 nm thickness of the QL  (inset of Fig.\ref{fig:1}b), as also observed in the work of Bansal et. al. \cite{bansal2014}. For transport measurements, thicker Bi$_2$Se$_3$ films in 10 to 16 nm range were grown. Subsequently, a 3 nm film of EuS was deposited in a sulphur-rich environment at room temperature, without breaking vacuum, followed by a 2 nm-capping layer of AlO$_x$(x$\sim$1.3-1.5). Hall bar structures with length, L and width, W were patterned by mechanical etching (Fig. 1a, also see Fig. S2). Compared to the Hall mobility ($\mu$) of epitaxial films of similar thickness grown on c-cut sapphire substrate ($\sim$1000 cm$^2$/V$\cdot$s for 16 nm), mobility of films on a-SiO$_2$ substrates were found to be lower ($\sim$450 cm$^2$/V$\cdot$s). This relatively lower bulk mobility of films on a-SiO$_2$ substrate may be beneficial in detecting a larger contribution of the robust SSs, unaffected by defects, over bulk conduction. 

\par Figure \ref{fig:1}b shows the temperature dependence of longitudinal resistivity ($\rho_{xx}$) for 12 nm Bi$_2$Se$_3$ films with three different capping layers viz. 2 nm AlO$_x$, 2 nm Se and 3 nm EuS. All the three devices show a predominant metallic signature during cool down, but respond differently at lower temperatures ($<30K$). Films with AlO$_x$ capping layer show a weak upturn in $\rho_{xx}$ at $\sim$28K followed by a metallic signature at lower temperatures ($<$19K). On the other hand, Bi$_2$Se$_3$ films with Se and EuS capping layer show an upward trend in $\rho_{xx}$ below 19K, with a stronger response in the latter. The above observation of capping layer dependent variations in the temperature dependence of $\rho_{xx}$, primarily in a narrow thickness window of $\sim$12 to 16 nm of our Bi$_2$Se$_3$ films, does suggest combination of bulk impurity band freezing \cite{analytis2010} and complex interface processes such as charge transfer across the interface due to work function mismatch/chemical reactivity and charge transfer between the TSSs and bulk states, all of them to be playing a role in the device response (see section 3 $\&$ 4 of Suppl. Mater.). In the case of TIs, position of E$_F$ at the surface can greatly depend on the relative concentration of bulk and surface defect density \cite{brahlek2015,veyrat2015}. A large mismatch in the surface and bulk E$_F$ can lead to charge transfer and bending of bulk bands near the surface, resulting in a depleted (an accumulated) surface region in films with lower (higher) surface defect density (see section 4 of Suppl. Mater.). Importantly, the decrease in $\rho_{xx}$ in AlO$_x$ capped films below $\sim$19K does suggest that apart from the insulating-like response due to impurity band freezing at $\sim$28K, interface band bending leading to accumulation is also playing a contributory role to the overall device resistance (see Fig S5). 

\par To investigate the upturn in film's $\rho_{xx}$ for the Se and EuS capped films, thickness dependence of $\rho_{xx}$ vs T plot is shown in figure \ref{fig:1}c. Hall mobility of these devices increase with cool down and finally saturate (Fig. \ref{fig:1}d), suggesting a dominant band-transport conduction (see section 3 of Suppl. Mater.). In the case of Se capped films, the upturn in $\rho_{xx}$ occur at $\sim$18K for 12 nm films and decrease to $\sim$15K for 16 nm films which we attribute primarily due to impurity band freezing. In the case of films of similar thickness and mobility, capped wih EuS, a stronger upturn in $\rho_{xx}$ occur at a temperature of 19K, that remain unchanged in both the 12 nm and 16 nm Bi$_2$Se$_3$/EuS devices. These observations suggest the origin of this upturn to a more dominant interface effect supporting an MIT response. Further, the decrease in $\rho_{xx}$ of the EuS capped devices with cool down is relatively gradual than the Se capped devices (Fig. \ref{fig:1}c), a trend observed in different samples with different W (200 $\mu$m to 2 mm). This response is primarily caused by depletion of charge carriers, noted by the relatively lower Hall carrier density ($n_{Hall}$, Fig. \ref{fig:1}e), that suggest contributions from the upward band bending (depletion region) near the Bi$_2$Se$_3$/EuS interface with the possible lowering of interface $E_F$ below the conduction band edge. It is experimentally known that the surface of Bi$_2$Se$_3$ can be chemically reactive that can cause strong band-bending effects at the surface \cite{benia2011,bianchi2010}. Density functional theory (DFT) studies reveal that Eu-Se bond formation at the Bi$_2$Se$_3$/EuS interface is more favorable \cite{lee2014} and hence may cause lower surface defect density supporting an upward band bending. Further studies may be needed to microscopically probe these interface chemical responses. Additionally, we also explored the effect of above insulating transition on the topological properties of Bi$_2$Se$_3$ SSs using weak antilocalization (WAL) studies \cite{brahlek2015,brahlek2014} at low temperatures ($<30$K, Fig. S7). Our analysis using the Hikami-Larkin-Nagaoka (HLN) formalism \cite{hikami1980} reveal a drop in WAL channel by a value of $\sim$1 in the Bi$_2$Se$_3$/EuS devices compared to the Bi$_2$Se$_3$/Se devices (showing a value close to 2). This finding corroborates well with the possibility of MPE opening up an EG in the top SS, with only the bottom surface channel dominantly contribute to the WAL response.   
\begin{figure}[t]
    \centering
    \includegraphics[width=\linewidth]{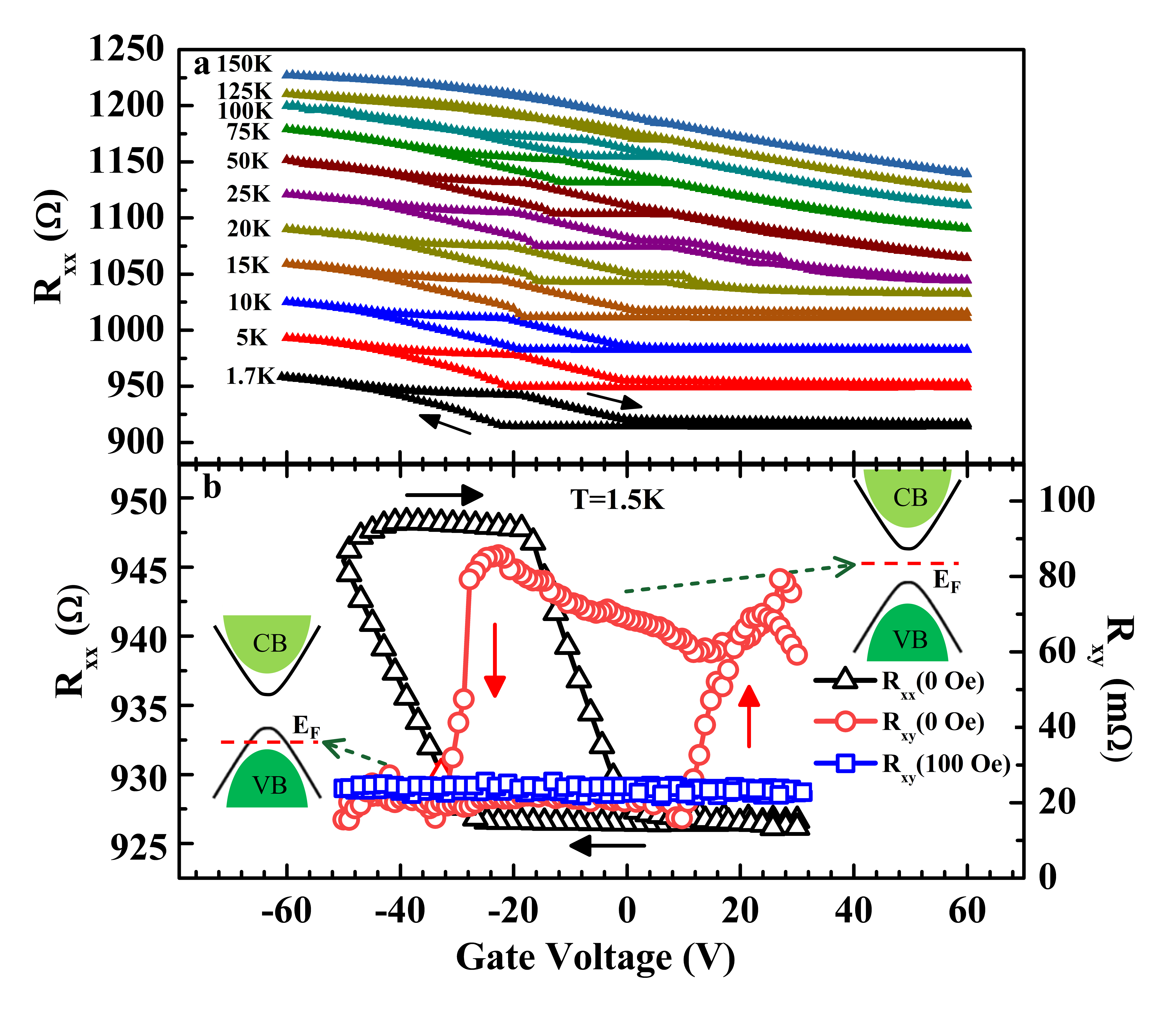}
    \caption{(a) $R_{xx}$ vs $V_g$ of the Bi$_2$Se$_3$(16 nm)/EuS(3 nm) device measured at different $T$s. A second hysteresis response appear in positive $V_g$ at and below 25K (refer to Fig. S6). The plots are shifted vertically for clarity. (b) $R_{xx}$ (triangles, left y-axis) and $R_{xy}$ (circle and square, right y-axis) vs $V_g$ for the above device at 1.5K. Red circle is measured at 0 Oe and blue square is taken in a perpendicular field of 100 Oe. The solid arrow heads denote the direction of sweep in $V_g$. Position of the $E_F$ corresponding to the high and low value of $R_{xy}$ (for 0 Oe data) is schematically shown.}
    
    \label{fig:2}
\end{figure} 
\par In an effort to tune E$_F$ within the bandgap to access the exchange gapped TSSs, the devices were investigated using back gate voltage ($V_g$) studies. Here, the high bulk carrier density ($>10^{19}$/cm$^3$) prevent the movement of bulk $E_F$ towards the Dirac point and is expected to only modulate the band bending near the depleted Bi$_2$Se$_3$/EuS interface. Figure \ref{fig:2}a shows the temperature dependence of longitudinal resistance ($R_{xx}$) vs $V_g$ of the Bi$_2$Se$_3$/EuS devices. Here, a hysteresis response in $R_{xx}$ is observed at all temperatures below 125K, with similar response observed in 4 different sets of samples and with different film thickness (Fig. S10). In comparison, the Se capped devices show no such behavior (Fig. S9). Our analysis suggest that the hysteresis in $R_{xx}$ arise due to weaker gating, with the interface E$_F$ pinning/de-pinning within the band gap of Bi$_2$Se$_3$ (primarily at the band edge). The fact that the onset of hysteresis occurs around the same temperature when the carrier density falls down ($T<100$K, Fig. \ref{fig:1}e) corroborates well with the mechanism related to band bending and the E$_F$ at the interface dropping into the band gap. In figure \ref{fig:2}a, at temperatures below 25K, a second smaller hysteresis response is noted, extending to positive gate voltage (refer to Fig. S10). We attribute the origin of the second hysteresis to the opening of EG by EuS and the associated E$_F$ pinning/de-pinning within the EG. In figure \ref{fig:2}b, we show $R_{xx}$ and Hall resistance ($R_{xy}$) vs $V_g$ of the device measured at 1.5K. $R_{xy}$, measured in zero magnetic field, also shows a hysteresis response with the value rising above $V_g$ $\sim$15V in the positive sweep cycle reaching $\sim$70m$\Omega$ and dropping down at -27V in the negative sweep cycle of $V_g$. We associate the high resistance state of $R_{xy}$ to the condition of $E_F$ lying in the EG of the SSs showing signature of H-IQHE (Fig. \ref{fig:1}a). Interestingly, as expected, the H-IQHE disappear in presence of a small perpendicular magnetic field of 100 Oe that aligns the domains in the EuS, without an appreciable change in $R_{xx}$. These observations further validate the existence of perpendicularly aligned magnetic domains that support dissipationless chiral edge states at the domain walls. Here, the existence of bulk conduction channel, multi-domain structure and bottom Dirac SS are limiting the observation of half-integer value of quantised Hall conductance.

\par
\begin{figure}[ht]
    \centering
    \includegraphics[width=\linewidth]{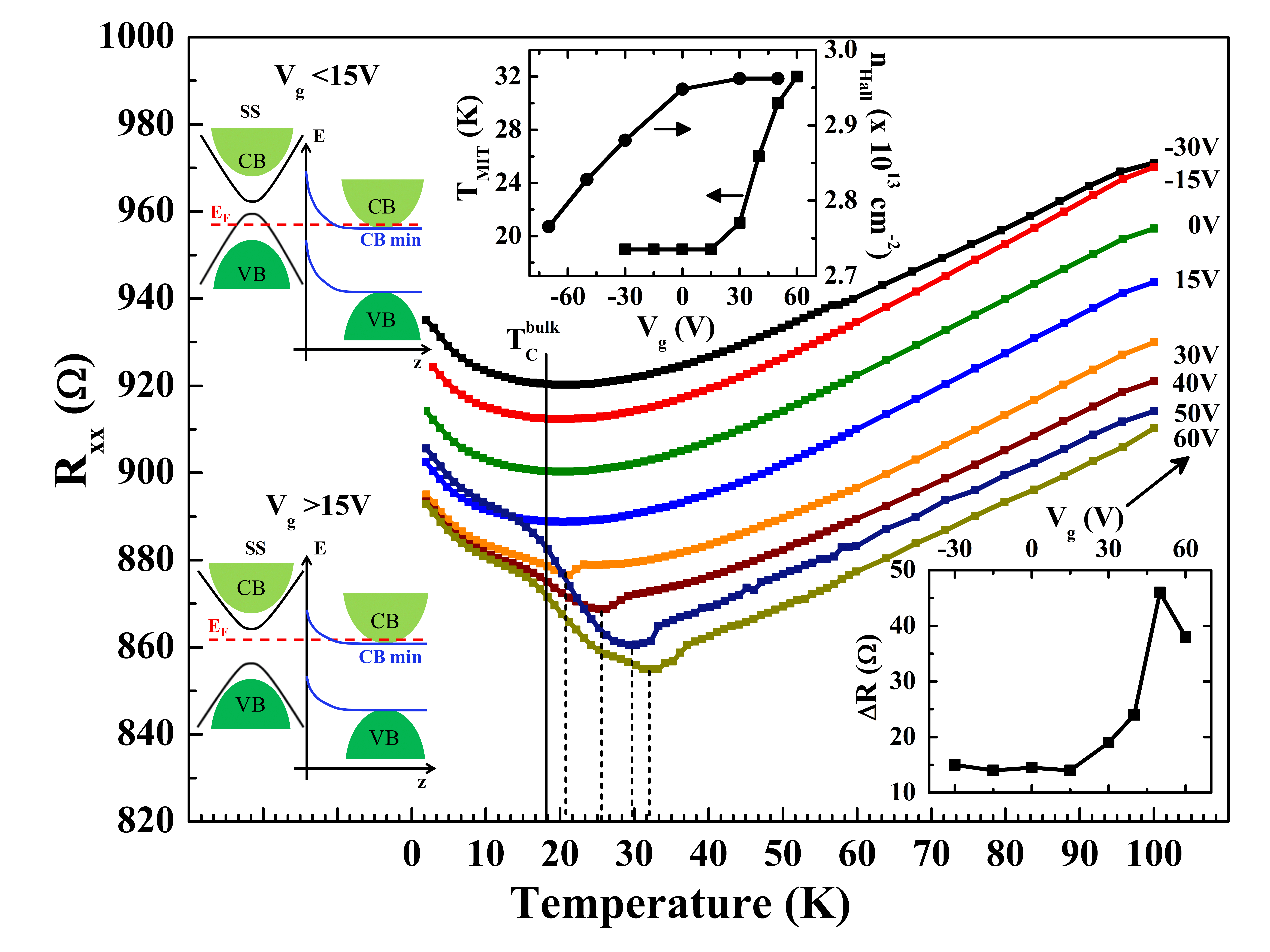}
    \caption{$R_{xx}$ vs $T$ of Bi$_2$Se$_3$(16 nm)/EuS(3 nm) device (L=1.5 mm, W=1 mm) measured at different $V_g$s. $T_c^{bulk}$ denotes the $T_c$ of the bulk EuS film, measured at negative $V_g$ (depleted carriers at interface). Top middle inset shows increase in  $T_{MIT}$ (square; shown by vertical lines in main figure) at positive $V_g$s which is corroborated by the increase in $n_{Hall}$ (circle). Left inset shows schematic of CB bending (solid lines) at the top Bi$_2$Se$_3$ surface for large positive $V_g$ (bottom) and large negative $V_g$ (top). Enhanced interface magnetism, attributed to indirect exchange at positive $V_g$s, can also lead to increased EG suggested by large $\Delta R=R(1.5K)-R(T_{MIT})$ peaking at $V_g=50$V (bottom right inset).}
    \label{fig:3}
\end{figure}

 We next probe the effect of $V_g$ on the MIT response, due to opening of EG, in our Bi$_2$Se$_3$/EuS devices. Figure \ref{fig:3} shows the measurements of $R_{xx}$ vs T, performed at different $V_g$s. At negative $V_g$, the insulating transition was observed at temperature, T$_{MIT}\sim19$K, which is enhanced to $\sim32$K at positive $V_g$ with an corresponding increase in the carrier density (top inset of Fig.\ref{fig:3}). In comparison, the Se capped devices showed no such response (see Fig. S9). Such an increase in $T_{MIT}$ at positive $V_g$, which also makes the device conducting (seen by the downward shift of plots), at first seems counter-intuitive. However, this observation is suggestive of conduction-electron-mediated ferromagnetic exchange between the Eu atoms strengthening the interface magnetism, with an enhancement in $T_c$ of the EuS 'interface layer'. In EuS, magnetism arises due to the overlap of the Eu $4f$ orbitals with the valence band below the E$_F$ leading to a weaker exchange coupling \cite{coey2009}. DFT studies by Kim et. al. \cite{kim2017} have shown that the TSSs that weakly interconnect the conduction and valance bands can mediate exchange coupling between Eu atoms through the Ruderman-Kittel-Kasuya-Yosida (RKKY) type interactions \cite{ruderman1954,kasuya1956,yosida1957}, with the exchange constant being sensitive to the position of E$_F$ near the Dirac point. The observed gate controlled tuning of $T_{MIT}$ in our devices does provide evidence of such an indirect exchange mechanism at the Bi$_2$Se$_3$/EuS interface. This enhanced interface magnetism can hence open up a larger EG and may be responsible for the larger increase in resistance under positive $V_g$, peaking at $V_g\sim50$V (bottom-right inset of Fig.\ref{fig:3}). A higher value of $T_{MIT}$ $\sim19$K i.e. 2K above the bulk $T_c$ of EuS is observed even at high negative $V_g$ (depleted interface) which may indicate the role of high spin-orbit strength of the Bi$_2$Se$_3$ surface states leading to an enhancement in the Tc of the bulk EuS film. For $V_g$ $>15$V, the resistance plots show a point of inflection at 19K, below which, the resistance increases further. This feature may therefore correspond to the magnetic transition of EuS layers away from the interface, unaffected by the above carrier mediated indirect exchange interactions. Comparing with the warm up data of $R_{xx}$ vs T  (Fig. S11), our analysis support the existence of difference in the transition temperature of bulk and interface EuS layer. 

\par 

\begin{figure}[t]
    \centering
    \includegraphics[width=\linewidth]{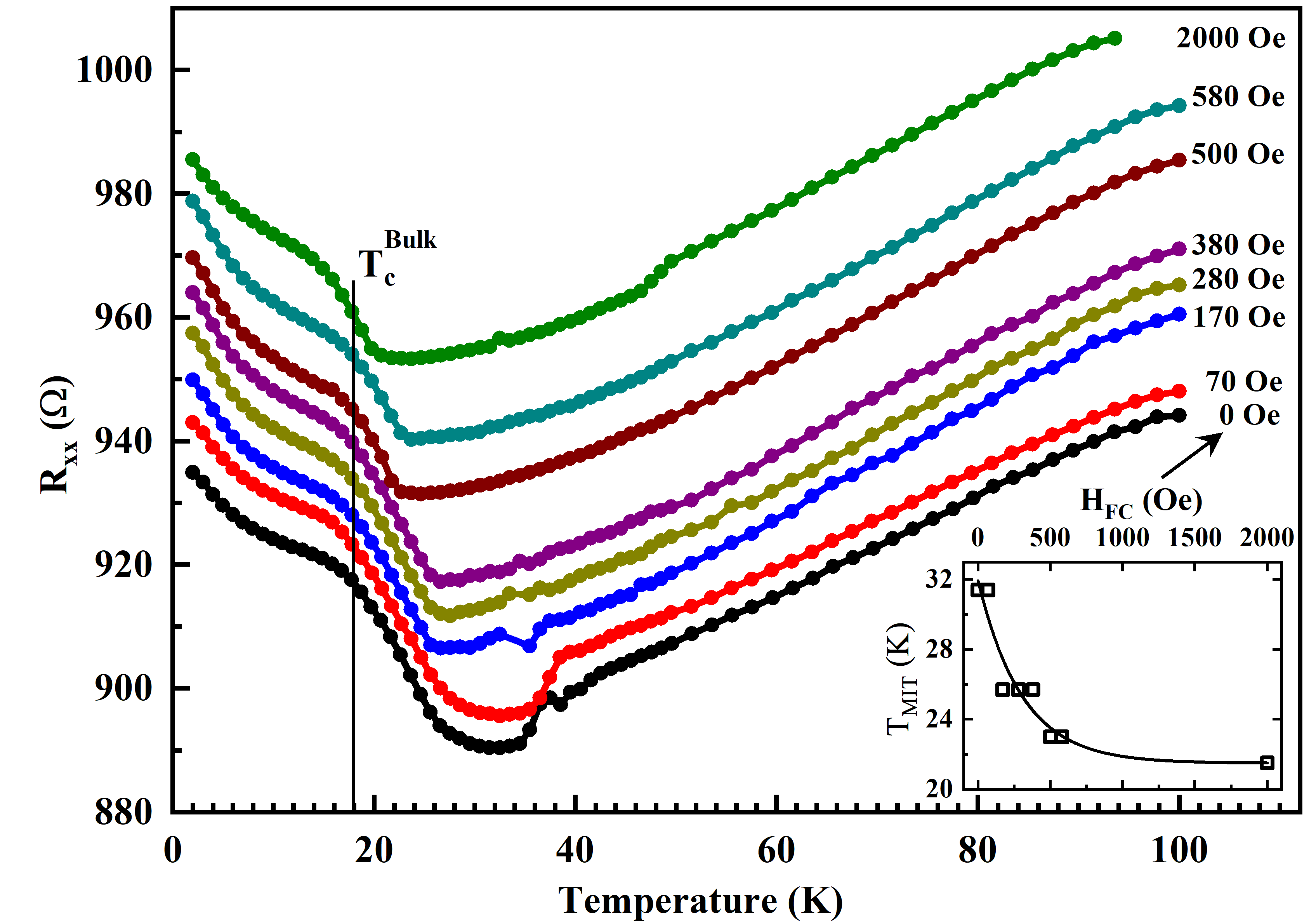}
    \caption{$R_{xx}$ vs $T$ of the Bi$_2$Se$_3$(16 nm)/EuS(3 nm) device measured at $V_g=50$V with varying perpendicular $H_{FC}$s. The plots are shifted vertically for clarity. Bottom right inset shows lowering of $T_{MIT}$ with $H_{FC}$ attributed to the weakening of the indirect exchange at the interface.}
    \label{fig:4}
\end{figure}

\par Apart from the enhancement in $T_{MIT}$ for $V_g$ $>$ 15V, we also observe a very prominent and reproducible drop (observed in different devices) in the device resistance preceding MIT. This sudden drop in resistance, by almost 10 $\Omega$ (for the plot of $V_g$=50V), may be suggestive of formation of 1D chiral fermion edge states \cite{hasan2010} along the magnetic domain walls at the interface around $T_c$ (Fig. \ref{fig:1}a), with the simultaneous opening of EG. With the lowering of temperature, domains become larger leading to a dominant insulating transport response due to EG opening. Comparing with the $R_{xy}$ vs $V_g$ data of figure \ref{fig:2}b (in the positive sweep cycle), the observation of the resistance drop for $V_g$ $>$ 15V in figure \ref{fig:3} corroborates well with the position of $E_F$ within the EG to start observing the H-IQHE signal and to access these chiral edge states in our devices. 
\par To further investigate these responses, the devices were cooled down in different perpendicular magnetic fields ($H_{FC}$), keeping $V_g$ at 50V, as shown in figure \ref{fig:4}. Firstly, we observe a decrease in $T_{MIT}$ with the increasing strength of $H_{FC}$ (inset of Fig.\ref{fig:4}). This may be suggestive of weakening of the antiferromagnetic exchange between the conduction electrons and the Eu $4f$ atoms, observed in such indirect exchange interactions. Also notably, when cooled in higher magnetic fields, the resistance drop preceding $T_{MIT}$ starts to vanish in these devices. This occurs because the domain with parallel magnetization grows at the onset of magnetism, resulting in the reduction of domain wall that support these conducting 1D edge states. These observations of the dip in resistance before MIT, which we attribute to the formation of conducting 1D edge states at the domain walls, also highlight that the interface-gapped states in our devices do carry topological properties, and the MIT response of our devices is not due to the formation of any trivial hybridization gap \cite{eremeev2015, dzero2016}. Together with the evidence of H-IQHE signal, our results therefore demonstrate the capability to use MPE to engineer new classes of topological electronic devices.

\par KVR acknowledges Ajit C. Balram for useful discussions. All authors acknowledge TIFR-Hyderabad, SERB Ramanujan, Department of Atomic Energy (DAE) and ECR grant for funding. SM acknowledges SERB NPDF fellowship. 


\providecommand{\noopsort}[1]{}\providecommand{\singleletter}[1]{#1}%
%

\onecolumngrid
\end{document}